\documentclass[a4paper,twoside]{article}
\newcommand{\affil}[1]{$^{\rm #1}$}
\usepackage[authoryear]{natbib}
\bibpunct{(}{)}{;}{a}{}{,}
\usepackage{graphicx,aastex}
\date{} 
\title{\large\bf\flushleft The Formation of Constellation III in the Large Magellanic Cloud}
\author{\parbox{\textwidth}{\flushleft
\vspace{-0.5cm}
{\it Jason Harris\affil{A,B} and Dennis Zaritsky\affil{A}}\\
\vspace{0.4cm}
{\small \affil{A}\,Steward Observatory, 933 N. Cherry Ave., Tucson, AZ 85721, USA}\\
{\small \affil{B}\,Email: jharris@as.arizona.edu}
}}
\begin{document}
\twocolumn[
\begin{changemargin}{.8cm}{.5cm}
\begin{minipage}{.9\textwidth}
\vspace{-1cm}
\maketitle
%
%
\small{\bf Abstract: We present a detailed reconstruction of the 
star-formation history of the Constellation~III region in the 
Large Magellanic Cloud, to constrain the formation 
mechanism of this enigmatic feature.  Star formation in 
Constellation~III seems to have taken place during two distinct 
epochs: there is the 8-15~Myr epoch that had previously been 
recognized, but we also see strong evidence for a separate 
``burst'' of star formation 25--30~Myr ago.  The 
``super-supernova'' or GRB blast wave model for the formation of 
Constellation~III is difficult to reconcile with such an extended, 
two-epoch star formation history, because the shock wave should 
have induced star formation throughout the structure 
simultaneously, and any unconsumed gas would quickly be dissipated, 
leaving nothing from which to form a subsequent burst of activity.
We propose a ``truly stochastic'' self-propagating star formation 
model, distinct from the canonical model in which star formation 
proceeds in a radially-directed wave from the center of 
Constellation~III to its perimeter.  As others have noted, and we 
now confirm, the bulk age gradients demanded by such a model are 
simply not present in Constellation~III.  In our scenario, the 
prestellar gas is somehow pushed into these large-scale arc 
structures, without simultaneously triggering immediate and violent 
star formation throughout the structure.  Rather, star formation 
proceeds in the arc according to the local physical conditions of the 
gas.  Self-propagating star formation is certainly possible, but in a 
truly stochastic manner, without a directed, large scale pattern.}



\medskip
\medskip
\end{minipage}
\end{changemargin}
]
\small

\section{Introduction}

Shapley's Constellation~III in the Large Magellanic Cloud (LMC) is one
of the most enigmatic structures in the local universe: a coherent
semicircular arc spanning several hundred parsecs, composed of
thousands of bright young stars and tens of star clusters.  Its 
regularity across such a large scale defies the fractal-like 
distributions that young stellar populations typically follow; in 
fact, Constellation~III may be unique in this regard.  In addition, 
Constellation~III is embedded inside the supergiant shell LMC~4, a 
circular hole in the LMC's HI disk that spans more than a kiloparsec 
\citep{kim99}, and whose rim is dotted with HII regions \citep{mea80}.

The singular nature of Constellation~III invites speculation about its
formation mechanism, which must have been similarly unique given the
absence of anything resembling this structure in other nearby galaxies.
\cite{wm66} popularized the ``Constellation~III'' designation, and
speculated that its stars were formed from material swept up in the
shock of a ``super-supernova''.  \cite{ee98} suggested that the
combined winds from a relatively small number of massive stars could
have swept LMC4 clean of gas, and subsequently triggered the formation
of Constellation~III.  However, as they point out in \cite{ee99}, the
unique nature of Constellation~III belies such a mundane formation
mechanism.  After all, the LMC is home to thousands of star clusters
which must have hosted similarly strong massive-star winds, yet it
contains no other structures like Constellation~III.  Instead,
\cite{ee99} favor an updated ``super-supernova'' idea, suggesting the
LMC4 cavity was blown by a gamma-ray burst formed by the coalescence
of an X-ray binary (perhaps ejected from the nearby rich star cluster
NGC 1978), and the swept-up material subsequently formed
Constellation~III.  \cite{dop85} presented the idea that
Constellation~III is the result of a stochastic self-propagating star
formation (SSPSF) process, directed in an outward radial propagation.
However, recent studies of the distribution of ages in
Constellation~III have not confirmed the radial age gradient reported
by \citeauthor{dop85} \citep{ols97, bra97, dh98}.

In this paper, we present a map of the past history of star formation
in the vicinity of Constellation~III, in order to constrain the
various formation scenarios for this unique structure.  In
Section~\ref{sec:data}, we briefly review the photometric data and our
StarFISH analysis software.  In Section~\ref{sec:map}, we present a
detailed map of the star formation history (SFH) throughout the
Constellation~III region.  We discuss the implications of our SFH map
in Section~\ref{sec:discuss}, and summarize the results in
Section~\ref{sec:summary}.

\clearpage

\begin{figure}[h]
\begin{center}
\includegraphics[scale=0.4, angle=0]{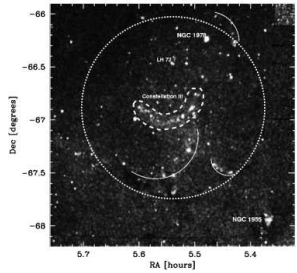}
\caption{A stellar flux density map of a $2.5^\circ\times2.5^\circ$
  region in the LMC, including Constellation~III.  The map was derived
  from our MCPS photometry: each pixel's value is proportional to the
  total stellar flux in $B$, $V$ and $I$ (for blue, green and red,
  respectively).  Major structures and clusters are labeled, including
  Constellation~III itself (dashed outline), and the approximate
  position of the LMC~4 supergiant shell (large dotted circle).  Note
  that Constellation~III is actually one of a few large stellar arcs
  in this region; our analysis will include all of these
  arcs. }\label{fig:constellationIII}
\end{center}
\end{figure}

\begin{figure}[h]
\begin{center}
\includegraphics[scale=0.4, angle=0]{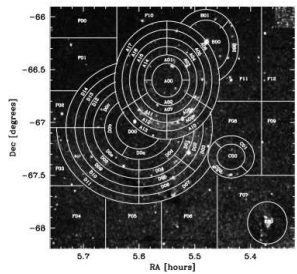}
\caption{The same flux-density maps as in
  Figure~\ref{fig:constellationIII}, with our conformal grid
  overplotted.  We determined the best-fit SFH solution independently
  for each grid cell in order to generate a map of the SFH in this
  region.}\label{fig:grid}
\end{center}
\end{figure}

\section{Overview of the Magellanic Clouds Photometric Survey and StarFISH}\label{sec:data}

The Magellanic Clouds Photometric Survey
\citep[MCPS,\ ][]{zar02,zar04} is a drift-scan survey of both the LMC
and the Small Magellanic Cloud (SMC), undertaken at the Las Campanans
Observatory 1-meter Swope telescope between 1995 and 2000.  The MCPS
provided CCD imaging to $V=21$~mag in $U$, $B$, $V$, and $I$ filters,
covering $8.5^\circ\times7.5^\circ$ in the LMC, and
$4^\circ\times4.6^\circ$ in the SMC.  Our catalogs contain astrometry
and photometry for 24~million LMC stars and more than 6~million SMC
stars.

In Figure~\ref{fig:constellationIII}, we show a stellar flux density
map derived from our MCPS photometry, for a $2.5^\circ\times2.5^\circ$
region including Constellation~III.  The Figure shows that the arc
traditionally known as Constellation~III is actually one of at least
three large stellar arcs in this region; all of these arcs lie in the 
interior of the LMC~4 HI supergiant shell.

\begin{figure}[h]
\begin{center}
\includegraphics[scale=0.4, angle=0]{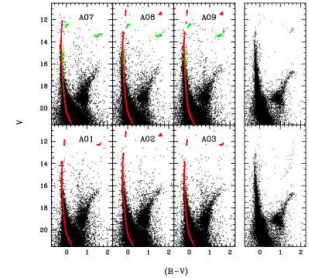}
\caption{The six panels on the left show $(B-V)$ vs. $V$ CMDs for six
  of our Constellation~III regions (as labeled at the top of each
  panel; see Figure~\ref{fig:grid}).  The termination of the main
  sequence at $V$=12--13~mag indicates that the youngest stars in
  these regions is aged 10--15~Myr (the red isochrones overplotted in
  each panel).  In addition, the three populations in the top row show
  an isolated clump of red supergiants near $V=13.5$~mag, indicative
  of a burst of star formation activity around 30~Myr ago (the green
  isochrones overplotted in the top panels).  This is further
  illustrated by the synthetic stellar populations in the rightmost
  column.  These two populations differ only in their SFH between 10
  and 100 Myr: in the top panel, the stars in this age range all have
  an age of 30~Myr, whereas in the bottom panel, the ages are
  uniformly distributed between 10 and 100~Myr.}\label{fig:cmds}
\end{center}
\end{figure}

The stellar populations of a galaxy represent a fossil record of its
past star-formation activity.  By statistically comparing multicolor
photometry of resolved stellar populations to synthetic populations
based on theoretical isochrones \citep[e.g., ][]{gir02}, we can
reconstruct the SFH of the target galaxy.  We have developed a
software package \citep[starFISH; ][]{hz01}, which performs robust SFH
analysis of resolved stellar photometry.  StarFISH works by
constructing a library of synthetic color-magnitude diagrams (CMDs),
each of which is built from isochrones spanning a small range in age
and metallicity.  The synthetic CMDs are designed to replicate the
characteristics of the observed data in every way (distance,
extinction, IMF, binarity, photometric errors, and incompleteness).
Each synthetic CMD therefore represents the contribution to the CMD of
stars of a particular age and metallicity.  Through linear combination
of synthetic CMDs spanning all relevant combinations of age and
metallicity, we can generate composite model CMDs that represent any
arbitrary SFH.  These composite model CMDs can then be quantitatively
compared to the real, observed CMD, and by minimizing the differences
between them, the best-fitting SFH solution can be obtained.

\section{Mapping the SFH of the Constellation~III Region}\label{sec:map}

In order to distinguish between the various competing theories for the
formation of Constellation~III, we constructed a spatially-resolved
map of the SFH of the entire $2.5^\circ\times2.5^\circ$ region.  In
order to maximize our ability to detect any radial or azimuthal
population gradients present in these stellar arcs, we constructed a
conformal grid that follows their curvature (see
Figure~\ref{fig:grid}), and determined an independent SFH solution for
the stars in each grid cell.  The synthetic CMDs employed for each
grid cell used extinction distributions and empirical photometric
error models derived directly from the grid cell's stellar population.

Previous analyses of the SFH of Constellation~III have largely sought
to determine a single characteristic age for different locations in
the structure, without considering the full distribution of stellar
ages that makes up the true SFH.  In particular, the ages assigned
have been the {\em youngest} age present.  As an illustration that a
more complete SFH is warranted for these regions, we examine the
$(B-V)$ CMDs of eight of our Constellation~III regions in
Figure~\ref{fig:cmds}.  Each of these regions shows a prominent main
sequence, and by simple isochrone fitting, one can determine the age
of the youngest stellar population in each region (as shown by the red
curves in each panel).  However, these CMDs show clear evidence for a
more complex SFH.  The red giant branch and red clump are obvious
tracers of old stellar populations, but there are also supergiants
present which trace star formation that occurred several tens of
millions of years ago.  In particular, the CMDs in the top row of
Figure~\ref{fig:cmds} each show an isolated knot of red supergiants
with $V=13.5$~mag.  An isolated knot of supergiants at a common
luminosity is strong evidence for an isolated burst of star formation
activity in the recent history of the region, because the luminosity
of a supergiant is directly and unambiguously correlated with its age
\citep{dp97}.  The isochrone overplotted in green in
Figure~\ref{fig:cmds} indicates that the supergiants in these knots
are roughly 30~Myr old.  Thus, by simple visual inspection of these
CMDs, we can already conclude that {\em some} of the Constellation~III
regions have experienced multiple, isolated bursts of star formation.
Variations like these will be recovered in our StarFISH analysis,
giving us a much more complete picture of the SFH of
Constellation~III.

\begin{figure*}[ht]
\begin{center}
\includegraphics[scale=0.8, angle=0]{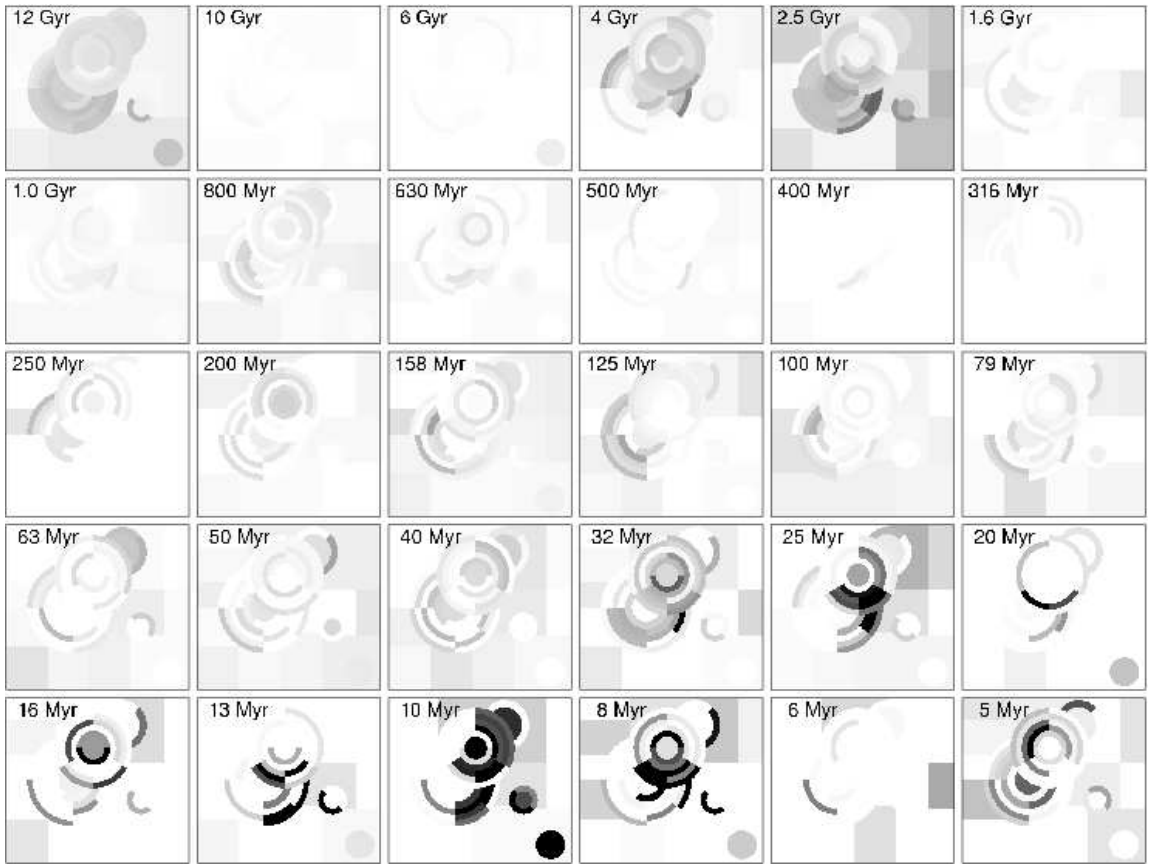}
\caption{The SFH map for the Constellation~III region.  Each panel
  displays the star formation rates for a single time step in the
  LMC's history, from 12~Gyr ago in the upper left to 5~Myr ago in the
  lower right.  Within a panel, the greyscale is proportional to the
  relative star formation rate in each grid cell (with darker color
  corresponding to a larger star formation rate).  }\label{fig:sfhmap}
\end{center}
\end{figure*}

\begin{figure}[h]
\begin{center}
\includegraphics[scale=0.3, angle=-90]{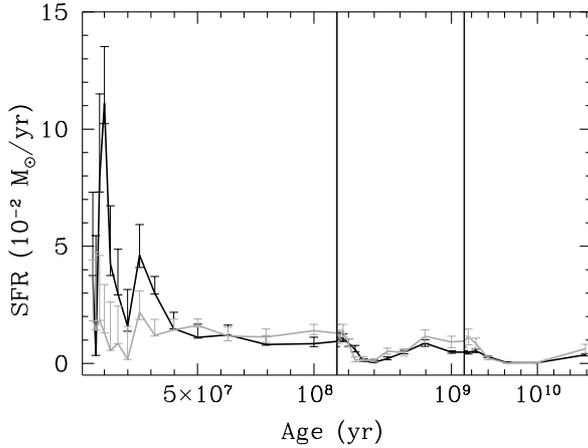}
\caption{The summed total SFH of the entire Constellation~III region,
  shown as the star formation rate as a function of time.  The time
  axis is displayed linearly, but the scale changes in the three
  panels so that the very narrow time intervals at the young end can
  be displayed.  The left panel covers about 100~Myr, the middle panel
  covers about 1~Gyr, and the right panel covers more than 10~Gyr.
  The black histogram represents the SFH of the stellar arcs in the
  Constellation~III region, while the grey histogram represents the
  SFH of the background population in this region.}\label{fig:totsfh}
\end{center}
\end{figure}

The SFH map of Constellation~III resulting from our StarFISH analysis
is shown in Figure~\ref{fig:sfhmap}.  In this Figure, each panel
represents a map of the past star formation rate for a different time
step, from 12~Gyr ago to 5~Myr ago.  It is immediately apparent from
the SFH map that star formation occurred in Constellation~III over an
extended time interval, from about 30~Myr ago until 8~Myr ago.  Star
formation was active in different parts of Constellation~III at
different times; however, there do not appear to be systematic,
large-scale age gradients revealed in these maps.  We also note that
Constellation~III does not distinguish itself from the background
stellar population until the onset of recent star-formation 30~Myr
ago, so it is likely that 30~Myr ago marks the time of its initial
formation.  This can also be seen in Figure~\ref{fig:totsfh}, the
summed total SFH for the entire Constellation~II region

When examining Figure~\ref{fig:sfhmap}, it is important to understand
that SFH maps like this do not allow us to display information on the
uncertainties associated with the best-fit star formation rates (SFRs)
in each time step.  However, StarFISH {\it does} estimate these
uncertainties, and they do include covariance between adjacent age
bins.  So, for example, in the map there are many regions that have a
large SFR in the 10~Myr panel, but a very low SFR in the 8~Myr panel,
and vice versa.  The uncertainties computed by StarFISH for these bins
indicates that these variations are not significant; in other words,
it is clear from the fit that there was a large amount of star
formation activity 8--10~Myr ago in many of these regions, but in most
cases, the data do not allow us to distinguish 8~Myr old stars from
10~Myr old stars, and this non-uniqueness is reflected in the computed
uncertainties.

\section{Discussion and Implications}\label{sec:discuss}

Most previous analyses of Constellation~III's SFH have concluded that
the stellar arc is 10--15~Myr old.  We have shown that the {\em
  youngest} stars in these arc structures are 10--15~Myr old, but that
they also contain abundant populations as old as 30~Myr.  This extended 
epoch of star formation is difficult to reconcile with currently-proposed 
ideas about the formation mechanism of Constellation~III.  In the
``super-supernova'' or GRB shockwave scenario, star formation would be 
triggered throughout the arc structure on a short timescale, resulting 
in a small age spread among the stars in Constellation~III.  The SSPSF 
scenario predicts a more protracted epoch of star formation, but as it 
is usually discussed in the literature \citep{dop85, ols97}, the 
propagation wave is directed radially outward from the center of 
Constellation~III.  Our SFH map confirms that there are no large-scale 
age gradients in Constellation~III that would be required by this model.

We propose a new scenario in which the pre-stellar material was swept
into large arcs (perhaps by dramatic forces such as a GRB-like
explosion, or the combined winds of massive stars), but that star
formation was not immediately triggered throughout the structure by
these forces.  Rather, star formation proceeded stochastically
throughout the giant prestellar cloud complex, according to the local
physical conditions of the interstellar medium.  Self-propagation may 
well be a part of this process.  We do not reject SSPSF {\em per se} in 
the formation of Constellation~III, but the large-scale 
radially-directed form in which it is usually discussed for this region.

It may seem unlikely that it would be possible to sweep material into
large, coherent structures without triggering massive, rapid star
formation in the material.  However, the LMC currently contains an
example of just such a large, coherent prestellar cloud complex: there
is a ridge of molecular gas extending more than 1.5~kpc southward from
30~Doradus, with a typical width of 100~pc \citep{miz01}.  This ridge
contains abundant molecular and atomic gas, and yet its specific star
formation rate is currently quite low.  While the physical processes
that led to the formation of this giant molecular ridge may be quite
different from the forces that scuplted Constellation~III, its
existence is evidence that it is at least possible to gather
prestellar material into a large coherent structure, without
simultaneously triggering rapid star formation throughout it.

\section{Summary}\label{sec:summary}

We present a reconstruction of the spatially-resolved SFH of the
enigmatic Constellation~III region in the northern LMC disk.  We find
that stars in the giant stellar arcs in this region formed over an
extended period, from 30 to 10~Myr ago.  While there are significant
spatial variations in the SFH, there don't appear to be large-scale
age gradients as would be expected in a SSPSF formation scenario.

Since our detailed SFH reconstruction of Constellation~III fits
neither of the widely-discussed formation scenarios for this unique
structure, we propose a new scenario in which the prestellar material
was swept up into large arcs, but star formation was not immediately
triggered throughout the cloud, or at least not violently so.  The
molecular ridge south of 30~Doradus provides evidence that it is
possible to organize kpc-scale coherent structures in prestellar
material without immediately triggering rapid star formation in the
gas.

\section*{Acknowledgments} 
Much of this work was performed while JH was supported by NASA
through Hubble Fellowship grant HF-01160.01-A awarded by the Space
Telescope Science Institute, which is operated by the Association of
Universities for Research in Astronomy, Inc., under NASA contract NAS
5-26555.  DZ acknowledges financial support from National Science 
Foundation grant AST-0307482 and a fellowship from the David and 
Lucille Packard Foundation.

\bibliographystyle{apj}
\bibliography{jharris}


\end{document}